\documentclass[conference]{IEEEtran}
\IEEEoverridecommandlockouts
\usepackage{amsmath}
\usepackage{amsfonts}
\usepackage{bbding}
\usepackage{amssymb}
\usepackage{array}
\usepackage{subfigure}
\usepackage{graphicx}
\usepackage{subfigure}
\usepackage[named]{algo}
\usepackage{psfrag}
\usepackage{xfrac}
\usepackage{booktabs}
\usepackage{stfloats}
\usepackage{cases}
\usepackage[compress]{cite}
\usepackage{hyperref}
\usepackage{bm}
\usepackage{multirow}
\usepackage{amsthm}
\usepackage{stfloats}
\usepackage{setspace}
\usepackage{color}
\usepackage{algorithm}
\usepackage{algpseudocode}
\usepackage{amsmath}
\usepackage{cleveref}

\makeatletter
\renewcommand{\citepunct}{,\penalty\@m\hskip.13emplus.1emminus.1em}
\renewcommand{\citedash}{\hbox{--}\penalty\@m}

\makeatother

\allowdisplaybreaks

\begin{document}
\vspace{-8mm}
\title{Proactive Optimization with Machine Learning: Femto-caching with Future Content Popularity}
\author{
\IEEEauthorblockN{{Jiajun Wu, Chengjian Sun and Chenyang Yang}}
\IEEEauthorblockA{Beihang University, Beijing, China\\Email: \{jiajunwu,sunchengjian,cyyang\}@buaa.edu.cn}
}

\maketitle
\begin{abstract}
Optimizing resource allocation with predicted information has shown promising gain in boosting network performance and improving user experience. Earlier research efforts focus on optimizing proactive policies under the assumption of knowing the future information. Recently, various techniques have been proposed to predict the required information, and the prediction results were then treated as the true value in the optimization, i.e., ``first-predict-then-optimize''.
In this paper, we introduce a proactive optimization framework for anticipatory resource allocation, where the future information is implicitly predicted under the same objective with the policy optimization in a single step. An optimization problem is formulated to integrate the implicit prediction and the policy optimization, based on the conditional distribution of the future information given the historical observations. To solve such a problem, we transform it equivalently to a problem depending on the joint distribution of future and historical information. Then, we resort to unsupervised learning with neural networks to learn the proactive policy as a function of the past observations via stochastic optimization.
We take proactive caching and bandwidth allocation at base stations as a concrete example, where the objective function is the conditional expectation of successful offloading probability taken over the future popularity given the historically observed popularity. We use simulation to validate the proposed framework and compare it with the ``first-predict-then-optimize'' strategy and a heuristic ``end-to-end'' optimization strategy with supervised learning.
\end{abstract}
\begin{IEEEkeywords}
Proactive optimization, future information, machine learning, femto-caching
\end{IEEEkeywords}

\vspace{-2mm}

\section{Introduction}
Anticipatory resource management can adaptive to dynamic user behavior or network environment in a proactive manner, which is an emerging technique in facing the unprecedented challenges in the fifth generation and beyond wireless systems \cite{Ma2020ProactOpt}.

To demonstrate the potential in harnessing future information such as user location and content popularity, earlier works optimize proactive policies towards various objectives such as throughput and energy efficiency, under the assumption of perfect prediction. For example, the file popularity in the next cache-update duration was assumed known for proactive caching in \cite{femtocachingTIT,cachemethod,LDHNcache}, and the data rates or locations of a mobile user in the next tens of seconds were assumed known for predictive resource allocation in \cite{Vec2013Info,YCT2016TCOM,GJ_2018VTC}.

To achieve the promising gain of proactive policies in improving network performance and user experience, many techniques have been proposed for making the prediction. Most research efforts adopt the ``divide-and-conquer'' strategy, which treat the information prediction as an independent task of the policy optimization, and take the predicted information as the true value in the optimization \cite{Zeydan2016big,NB2018}.
Although viable, the prediction obtained from the loss function (say mean square error \cite{NB2018}) may not perform well in terms of the ultimate goal of the policy optimized towards another objective (say throughput). Moreover, whenever the concerned user behavior or network environment changes, the required information needs to be re-predicted and the proactive policy has to be re-optimized again.

In fact, the information prediction and the policy optimization can be obtained with a single objective in a single step, under a \emph{proactive optimization framework}. Since the prediction is made from the past data, the objective functions of such type of optimization depend on the conditional distribution of future information given historical observations.

In this paper, we establish a proactive optimization framework to formulate the prediction and optimization problem in an end-to-end manner, where the objective function is the conditional expectation of a metric taken over the future information given the past observations.
The key challenge of solving such a problem lies in the unknown conditional distribution, which is hard to estimate when the observation space is continuous. To circumvent such a difficulty, we transform the formulated variable optimization problem into an equivalent functional optimization problem, which relies on the joint distribution of the future information and the historical observation.
The transformed problem can be solved with an unsupervised learning framework designed for functional optimizations, where the joint distribution can be learnt through stochastic sampling \cite{SCJPIMRC}.

The basic principle of proactive optimization is similar to reinforcement learning, which is a model-free framework also providing implicit prediction \cite{Zhong2018deep,Liu2019GC}. Nonetheless, our framework can leverage the available models of wireless problems, in the form of the gradient of the metric with respect to the variables to be optimized.

To demonstrate how to formulate and solve a proactive optimization problem, we take proactive femto-caching as a concrete example. In particular, we jointly optimize the caching and bandwidth allocation policy at the base stations (BSs) according to the past content popularity to maximize the successful offloading probability (SOP). We use a real dataset to evaluate the proposed ``end-to-end'' optimization strategy by comparing with existing strategies.


\section{Proactive Optimization}
Consider a proactive optimization problem with metric function $J(\cdot)$ and constraint function $c(\cdot)$, which can be formulated in a genetic form as follows,
\begin{align}\label{pro-opt-1}\vspace{-1mm}
    \max\limits_{x^t}\ & \mathbb{E}_{f^t \mid h^{t-1}} \left\{J \left(f^t , x^t\right)\right\} \\
    {\rm s.t.}\ & c\left(x^t , h^{t-1}\right) \leq 0,\nonumber
\end{align}
where the variable $x^t$ is optimized and the future information $f^t$ is implicitly predicted both according to the historical observation $h^{t-1}$, $\mathbb{E}_{f^t \mid h^{t-1}} \left\{\cdot\right\}$ in the objective denotes the conditional expectation taken over $f^t$ given $h^{t-1}$.

In this generic formulation, $x^t$ may denote a single variable (say transmit power to a user) or multiple variables (say transmit powers to multiples users). $h^{t-1}$ or $f^t$ may correspond to a single user behavior or network environment parameter (say a channel gain) or multiple parameters (say a channel vector). $f^t$ may represent the information for the same type of parameter as $h^{t-1}$ (say $f^t$ is the popularity of a file in time period $t$ and $h^{t-1}$ represents the popularity of the file in previous time periods), such that $\{f^t, h^{t-1}\}$ is a time series (say dynamic popularity of the file). $f^t$ may also represent the parameter differing from but related to $h^{t-1}$ (say  $f^t$ is the average channel gain of a user in time period $t$ while $h^{t-1}$ represents the locations of the user in previous time periods).


Stochastic optimization is a powerful tool for finding the solution of a problem with unknown distribution via sampling the random variables, which however is not applicable to the problem in \eqref{pro-opt-1}. Different from the optimization that accounts for the uncertainty of future information but not exploiting the historical observations \cite{CY2019TCOM}, which requires the distribution of future information $\mathbb{P} \left\{f^t\right\}$, the objective in \eqref{pro-opt-1} depends on the conditional distribution $\mathbb{P} \left\{f^t \mid h^{t-1}\right\}$, which is hard to estimate in practice. This is because the historical observations are recorded from real environments, which cannot not be generated with unknown conditional distribution. For the observations of $\{f^t, h^{t-1}\}$ recorded in the past that are random sequences, one realization of $h^{t-1}$ corresponds to only one realization of $f^{t}$. Yet multiple realizations of $f^t$ are required for one realization of $h^{t-1}$ to learn the conditional distribution $\mathbb{P} \left\{f^t \mid h^{t-1}\right\}$. As a result, solving the proactive optimization problem is non-trivial.


Noticing the fact that the distribution $\mathbb{P} \left\{f^t, h^{t-1}\right\}$ can be estimated from multiple realizations of $\{f^t, h^{t-1}\}$, we transform the variable optimization problem in \eqref{pro-opt-1} into an equivalent functional  optimization problem. In particular, we further take the average of the objective in \eqref{pro-opt-1} over $h^{t-1}$ and optimize the relation between the concerned variable and the historical observations denoted, i.e.,
\begin{align}\label{pro-opt-2}
\max\limits_{\substack{x^t(h^{t-1})}}\ & \mathbb{E}_{h^{t-1}} \left\{ \mathbb{E}_{f^t \mid h^{t-1}} \left\{ {J \left(f^t,x^t(h^{t-1})\right)}\right\} \right\} \\
{\rm s.t.}\ & c\left(x^t(h^{t-1}), h^{t-1}\right) \leq 0, \nonumber
\end{align}
where $\mathbb{E}_{h^{t-1}} \left\{ \mathbb{E}_{f^t \mid h^{t-1}} \left\{\cdot\right\}\right\} = \mathbb{E}_{f^t, h^{t-1}} \left\{\cdot\right\}$ is the joint expectation taken over $\{{f}^t,{h^{t-1}}\}$.

The problem in \eqref{pro-opt-1} is equivalent to the problem in \eqref{pro-opt-2} in terms of yielding the same optimal solution, as proved in Appendix A in \cite{SCJPIMRC}.

The problem  in \eqref{pro-opt-2} is a functional optimization where the ``variable'' to be optimized is a function. This type of problems cannot be solved with standard tools such as interior point method, even if the objective function can be derived with closed form for a known joint distribution. To find the solution of this problem, we resort to a framework of unsupervised learning  with deep neural networks (DNNs) proposed in \cite{Mark2019Sig, SCJPIMRC}, where stochastic optimization is used to cope with the unknown joint distribution $\mathbb{P} \{ {f}^t, {h}^{t-1} \}$.


\section{A Concrete Example: Proactive Femto-caching Optimization}
Consider a cellular network, where the BSs and users are located following two independent Poisson point processes with intensity $\lambda_b$ and $\lambda_u$, respectively. Each BS with a single antenna is connected to the core network via backhaul, and can cache $C$ files. Each user with a single antenna requests files from a content library with $F$ files.

Time is discretized into period each with cache update duration. Denote the file popularity in the $t$th time period as ${\bf p}^t=[p_1^t,...,p_{F}^t]$, where $p_f^t \in [0,1]$ is the probability that the $f$th file is requested by all users in the period satisfying $\sum_{f=1}^{F} p_f^t=1$.
Consider a probabilistic caching policy ${\bf q}^t=[q_1^t,...,q_{F}^t]$, where $q_f^t \in [0,1]$ is the probability that the $f$th file is cached at each BS in the $t$th time period satisfying $\sum_{f=1}^{F} q_f^t \leq C$. Once ${\bf q}^t$ is determined, the files are cached according to the method proposed in \cite{cachemethod}.

The user requesting the $f$th file is associated to the nearest BS that caches the $f$th file. During off-peak time, some BSs may not be associated with users, which are muted. Denote the probability of each BS being active as $p_{\rm a}$.

To avoid inter-cell interference, consider a random bandwidth allocation policy, where the total bandwidth $W$ of the network is divided into $I^t$ subbands, where $I^t$ is an integer. In  the $t$th time period, each BS transmits over a randomly selected subband occupying $\beta^t W$ bandwidth, where $\beta^t=1/I^t$ is a bandwidth allocation factor. Each BS has transmit power $P$. When a BS is associated with more than one user, the BS serves the users with frequency division multiple access, where the power and the bandwidth of the BS are equally allocated to each associated user.
For a user requesting the $f$th file and being associated with BS $b_0$ serving $U_f^t$ users, the received signal-to-interference ratio is
$\gamma_f^t \!=\! \frac {(P/U_f^t) g_0 r_0^{-\alpha}} {\sum\nolimits_{i \in \Phi_{b_0}^t} (P/U_f^t) g_i r_i^{-\alpha}}  = \frac {g_0 r_0^{-\alpha}} {\sum\nolimits_{i \in \Phi_{b_0}^t} g_i r_i^{-\alpha}}$,
where $g_0$ and $g_i$ are the channel gains from the user to the associated BS and other BSs, respectively, $r_0$ and $r_i$ are the corresponding distances, $\alpha$ is the path-loss coefficient, and $\Phi_{b_0}^t$ is the set of the other active BSs sharing the same subband with BS $b_0$. Then, the achievable data rate is $R_f^t = \frac{\beta^t W}{U_f^t} \log_2 \left(1+\gamma_f^t\right)$.

\subsection{Optimization with Known Future Popularity}
We consider a joint bandwidth and caching resource optimization problem to maximize the SOP of this example system.
SOP is defined as the probability that a user can be served with a data rate higher than a threshold $R_0$ by a BS that caches the requested file, which can be expressed as,
\begin{equation}
\begin{split}
p_{\rm s}( {\bf p}^t,\! \beta^t,\! {\bf q}^t) \triangleq \sum\limits_{f=1}^{F}\sum\limits_{n=1}^{\infty}p_f^t \Pr\left\{U_f^t = n\right\} \Pr\left\{R_f^t \!\geq\! R_0\right\},\nonumber
\end{split}
\end{equation}
where $\Pr\{\cdot\}$ denotes the probability.
According to Proposition 1 in \cite{LDHNcache}, the SOP can be approximated as
\begin{equation} \label{ps_ini}
\begin{split}
p_{\rm s}( {\bf p}^t \!,\! \beta^t \!,\! {\bf q}^t) \!\approx\! \sum_{f=1}^{F}   \frac{ {p_f^t q_f^t }} {q_f^t \!\!+\! p_{\rm a} \beta^t \!\left(\!q_f^t Z_{1,\gamma_{0,\beta}^t} \!\!\!+\! K \!\left(\!1 \!-\! q_f^t\!\right) \!\left(\!\gamma_{0,\beta}^t\!\right)^{\frac{2}{\alpha}}\!\right)},
\end{split}
\end{equation}
where $p_{\rm a} \approx 1- (1+\frac{\lambda_u}{3.5 \lambda_b})^{-3.5} $, $Z_{1,\gamma_{0,\beta}^t}=(\gamma_{0,\beta}^t)^{\frac{2}{\alpha}}$ $ \int_{(\gamma_{0,\beta}^t)^{-\frac{2}{\alpha}}}^\infty$ $ \frac{1}{1+x^{\frac{\alpha} {2} } }\,dx$, $K=\Gamma(1-\frac{2}{\alpha}) \Gamma(1+\frac{2}{\alpha})\Gamma(1)^{-1} $, $\gamma_{0,\beta}^t\approx 2^{\frac{R_0}{\beta^t W}(1+1.28 \frac{\lambda_u}{\lambda_b} )}-1$, and $\Gamma(\cdot)$ denotes the Gamma function. The approximation in \eqref{ps_ini} is accurate when $\frac{\lambda_u}{\lambda_b} \ll 1$ \cite{LDHNcache}.\footnote{Extensive simulation results show that the approximation in \eqref{ps_ini} is accurate even when $\frac{\lambda_u}{\lambda_b}=1$.}

When the popularity ${\bf p}^t$ is known in advance at the start of the $t$th time period, the bandwidth allocation factor $\beta^t$ and the caching policy ${\bf q}^t$  for the $t$th time period can be jointly optimized from the following optimization problem,
\vspace{-1mm}
\begin{subequations}\label{p_0}
\begin{align}
    {\bf P0}:\quad \max\limits_{\beta^t,{\bf q}^t}\ & p_{\rm s}\left({\bf p}^t,\beta^t, {\bf q}^t\right) \nonumber \\
    {\rm s.t.}\ & \sum_{f=1}^{F} q_{f}^t \leq C, \label{p1_a} \\
    & 0 < \beta^t \leq 1, \label{p1_b} \\
    & 0 \leq q_{f}^t \leq 1. \label{p1_c}
\end{align}
\end{subequations}

In practice, file popularity is unknown \emph{a priori}. To optimize proactive caching for the $t$th time period, ${\bf p}^t$ needs to be predicted from the historically observed popularities ${\bf h}^{t-1} \triangleq \left[{\bf p}^{t-1}, \cdots, {\bf p}^{t-\tau}\right]$ in an observation window composing $\tau$ time periods. To obtain viable policy for practical use, most existing works first predict the popularity and then treat the predicted popularity as the true value to solve ${\bf P0}$.

\subsection{Problem Formulation for Proactive Optimization}
Since ${\bf p}^t$ can be inferred from ${\bf h}^{t-1}$, we can allocate the caching and bandwidth resources at the start of the $t$th time period according to the past observations ${\bf h}^{t-1}$.
Under the proactive optimization framework, the optimization problem  can be formulated as follows,
\begin{subequations}\label{p_1}
\begin{align}
    {\bf P1}:\quad \max\limits_{\beta^t,{\bf q}^t}\ & \mathbb{E}_{{\bf p}^t \mid {\bf h}^{t-1}} \left\{p_{\rm s}\left({\bf p}^t,\beta^t, {\bf q}^t\right)\right\} \nonumber \\
    {\rm s.t.}\ & \eqref{p1_a}, \eqref{p1_b}, \eqref{p1_c}, \nonumber
\end{align}
\end{subequations}
where $\mathbb{E}_{{\bf p}^t \mid {\bf h}^{t-1}} \left\{\cdot\right\}$ denotes the conditional expectation taken over ${\bf p}^t$ given ${\bf h}^{t-1}$.

Solving problem ${\bf P1}$ is challenging, because the unknown conditional distribution $\mathbb{P} \{ {\bf p}^t | {\bf h}^{t-1} \}$ is hard to estimate. In the next subsection, we show how to convert ${\bf P1}$ into an equivalent form, which requires joint distribution. Then, we solve the equivalent functional optimization problem with unsupervised learning, which employs stochastic optimization to deal with the unknown joint distribution.

For notational simplicity, we use ${\bf h}$ to represent ${\bf h}^{t-1}$ in the rest of the paper unless otherwise specified.

\subsection{Proactive Caching and Bandwidth Optimization}

%

We convert ${\bf P1}$ into the following problem as we transform \eqref{pro-opt-1} into \eqref{pro-opt-2}, which finds the optimal policies as functions of the historical observation, $\beta^t({\bf h})$ and ${\bf q}^t({\bf h})$, to maximize the expectation of the objective function in ${\bf P1}$ over ${\bf h}$,
\begin{align*}
{\bf P2}: \max\limits_{\substack{\beta^t({\bf h}), {\bf q}^t({\bf h})}}\ & \mathbb{E}_{{\bf h}} \left\{ \mathbb{E}_{{\bf p}^t \mid {\bf h}} \left\{ {p_{\rm s}\left({\bf p}^t,\beta^t({\bf h}), {\bf q}^t({\bf h})\right)}\right\} \right\} \nonumber \\
{\rm s.t.}\ & \eqref{p1_a}, \eqref{p1_b}, \eqref{p1_c}, \nonumber
\end{align*}
where $\mathbb{E}_{{\bf h}} \left\{ \mathbb{E}_{{\bf p}^t \mid {\bf h}} \left\{\cdot\right\}\right\} = \mathbb{E}_{{\bf p}^t, {\bf h}} \left\{\cdot\right\}$, which is the joint expectation taken over $\left({\bf p}^t,{\bf h}\right)$ and can be learned via sampling.

Problem ${\bf P1}$ can be solved by solving problem ${\bf P2}$ via stochastic optimization.
To tackle with the constraints, we reconsider problem $\bf P2$ in its dual domain. The Lagrangian function of $\bf P2$ can be expressed as $\mathbb{E}_{{\bf p}^t,{\bf h}} \left\{L\left({\bf p}^t,\beta^t({\bf h}),{\bf q}^t({\bf h}),{\bm \xi}^t({\bf h})\right)\right\}$, where
\vspace{-1mm}
\begin{align}\label{Lagrangian}
    L &\left({\bf p}^t,\beta^t({\bf h}),{\bf q}^t({\bf h}),{\bm \xi}^t({\bf h})\right)\nonumber \\
    &\quad = p_s\left({\bf p}^t,\beta^t({\bf h}), {\bf q}^t({\bf h})\right) - \xi_c^t({\bf h}) \left(\sum_{f=1}^{F}{q_f^t({\bf h})} - C\right)\nonumber\\
    &\quad \quad - \sum_{f=1}^{F}\xi_f^t({\bf h}) \left(q_f^t({\bf h}) - 1\right),
\end{align}
$ {\bm \xi}^t({\bf h})=[\xi_c^t({\bf h}),{\bm \xi}_{f}^t({\bf h})]$ are the Lagrange multipliers, and ${\bm \xi}_{f}^t({\bf h})=[{\xi}_{ 1}^t({\bf h}),...,{\xi}_{F}^t({\bf h})]$.
Then, problem ${\bf P2}$ can be transformed into the following primal-dual problem,
\begin{align*}
{\bf P3}: \min\limits_{ {\bm \xi}^t({\bf h}) }  \max\limits_{\substack{ \beta^t({\bf h}), {\bf q}^t({\bf h}) } } \ & \mathbb{E}_{{\bf p}^t,{\bf h}} \left\{ L\left({\bf p}^t,\beta^t({\bf h}),{\bf q}^t({\bf h}),{\bm \xi}^t({\bf h})\right) \right\} \nonumber \\
{\rm s.t.}\ & \eqref{p1_b}, \xi_c^t({\bf h}), \xi_f^t({\bf h}), q_f^t({\bf h}) \geq 0.
\end{align*}

Problem $\bf P3$ is a functional optimization problem \cite{gregory2018constrained}, since the ``variables'' to be optimized are functions. To solve such a problem, we resort to the unsupervised learning framework in \cite{SCJPIMRC}. Specifically, we introduce four \emph{fully-connected} neural networks $ \tilde{\beta} ({\bf h};\theta_\beta)$,  $\tilde{\bf q}( {\bf h};\theta_{\bf q} )$, $\tilde{\xi}_c({\bf h};\theta_{ \xi_c} )$ and $\tilde{\bm \xi}_{f}({\bf h};\theta_{{\bm \xi}_{f}} )$ to approximate $\beta^t({\bf h})$, ${\bf q}^t({\bf h})$, ${\xi}_c^t({\bf h})$ and ${\bm \xi}_f^t({\bf h})$, where $\theta_\beta$, $\theta_{\bf q}$, $\theta_{\xi_c}$ and $\theta_{{\bm \xi}_{f}}$ are the model parameters of the neural networks, respectively.

To ensure \eqref{p1_b}, \texttt{Sigmoid} (i.e., $y=\frac{1}{1+e^{-x}}$) is used as the activation function for the output layer of $\tilde{\beta}({\bf h};\theta_{\beta})$.
To ensure $\xi_c^t({\bf h}), \xi_f^t({\bf h}), q_f^t({\bf h}) \geq 0$, \texttt{ReLU}  (i.e., $y=\max\{x,0\}$) is used as the activation function for the output layer of $\tilde{\xi}_c({\bf h};\theta_{ \xi_c} )$, $\tilde{\bm \xi}_{f}({\bf h};\theta_{{\bm \xi}_{f}} )$ and $\tilde{\bf q}({\bf h};\theta_{\bf q} )$.


Denote $\tilde{\bm \xi} ({\bf h};\theta_{\bm \xi})\triangleq$ $[\tilde{\xi}_c({\bf h};\theta_{ \xi_c} )$, $\tilde{\bm \xi}_{f}({\bf h};\theta_{{\bm \xi}_{f}} )]$, where $\theta_{\bm \xi}=$ $[\theta_{\xi_c},$$\theta_{{\bm \xi}_{f}}]$. By replacing $\beta^t({\bf h})$, ${\bf q}^t({\bf h})$ and ${\bm \xi}^t({\bf h})$ with  $\tilde{\beta} ({\bf h};\theta_\beta)$,  $\tilde{\bf q}( {\bf h};\theta_{\bf q} )$ and $\tilde{\bm \xi} ({\bf h};\theta_{\bm \xi})$, respectively, problem ${\bf P3}$ can be re-written as the following variable optimization problem,
\begin{equation*}
\label{p_4} {\bf P4}:  \min\limits_{\theta_{\bm \xi}} \max\limits_{ \theta_\beta ,\theta_{\bf q} } \  \mathbb{E}_{{\bf p}^t,{\bf h}} \left\{ L\left({\bf p}^t,\tilde{\beta} ({\bf h};\theta_\beta),\tilde{\bf q}( {\bf h};\theta_{\bf q} ),\tilde{\bm \xi} ({\bf h};\theta_{\bm \xi})\right)\right\}
\end{equation*}

To solve problem ${\bf P4}$, we can employ the primal-dual stochastic gradient method by using the objective function of ${\bf P4}$ as the loss function. In particular, the model parameters $(\theta_\beta,\theta_{\bf q})$ and $\theta_{\bm \xi}$ are updated along the ascent and descent directions of sample-averaged gradients, respectively.
The iterative formula is provided in Appendix \ref{App:Iter}, where a batch of samples are used for updating the DNN parameters in each iteration. By sampling $\left({\bf p}^t,{\bf h}\right)$ from the environment, the DNNs can learn the joint distribution $\mathbb{P}  \left\{{\bf p}^t,{\bf h}\right\}$. In this way, the proactive optimization problem in ${\bf P1}$ with implicit prediction is solved by finding the solution of ${\bf P4}$.

\begin{figure}[htbp]
    \centering
    \includegraphics[width=0.48\textwidth]{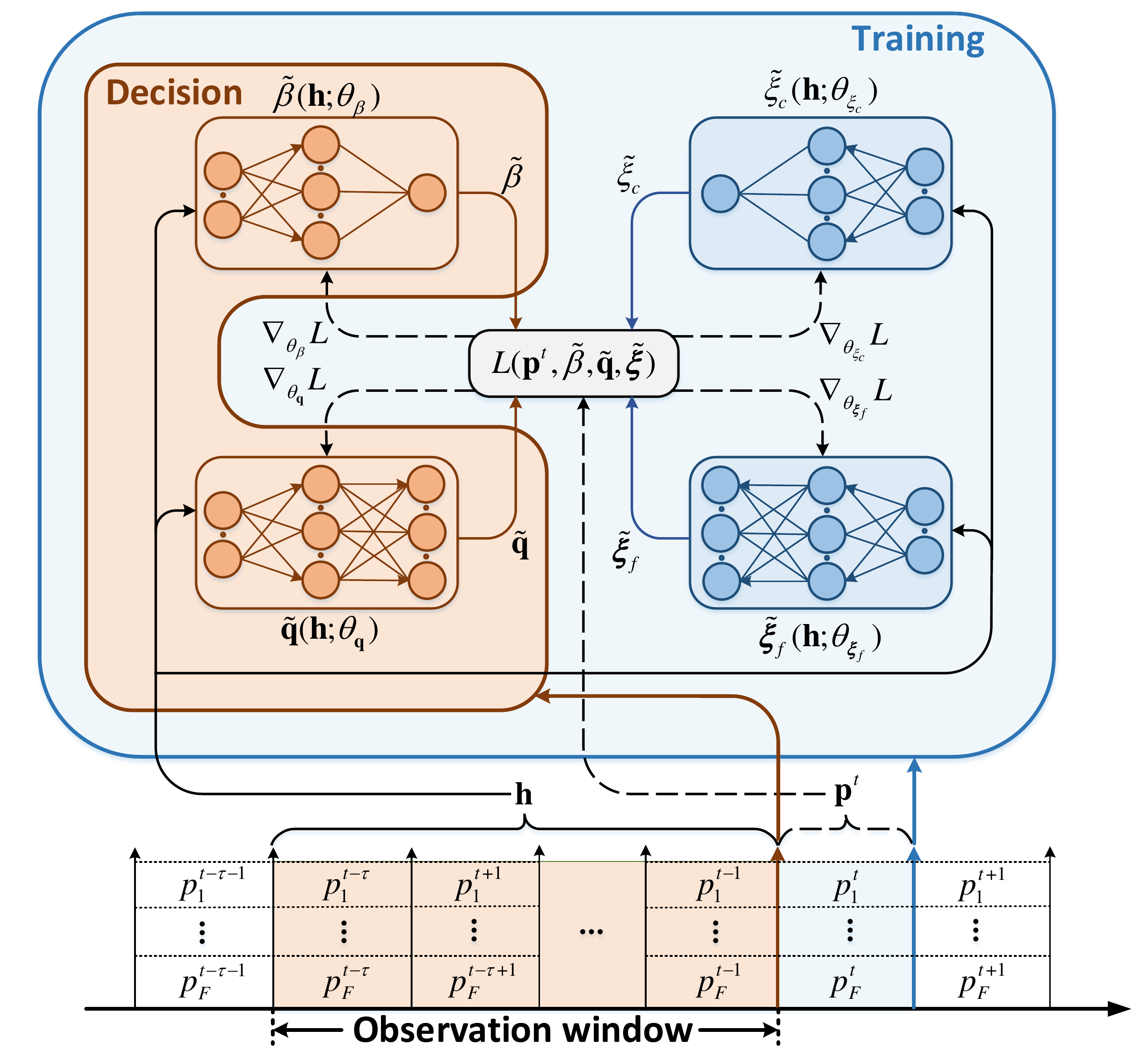}\vspace{-5mm}
    \caption{Decision and training for proactive optimization with DNNs.}
    \label{NN_fig}
\end{figure}

The DNNs used for making the decision for optimization is shown in Fig. \ref{NN_fig}. All the four DNNs use ${\bf h}$ (i.e., the historically observed popularity) as the input, and respectively output $ \tilde{\beta} ({\bf h};\theta_\beta)$,  $\tilde{\bf q}( {\bf h};\theta_{\bf q} )$, $\tilde{\xi}_c({\bf h};\theta_{ \xi_c} )$ and $\tilde{\bm \xi}_{f}({\bf h};\theta_{{\bm \xi}_{f}} )$, where the model parameters are trained via iterating the primal and dual variables provided in Appendix \ref{App:Iter} by using the Lagrangian function in \eqref{Lagrangian} as the loss function.
\vspace{-2mm}

\vspace{-2mm}
The DNNs can be trained either in an off-line or on-line manner. In the on-line operation, the batch size in Appendix \ref{App:Iter} is taken as one, i.e., $|\mathcal B| = 1$, and the observation in the ($t-1$)th period $\left({\bf p}^{(t-1)},{\bf h}^{(t-2)}\right)$ is used as the training sample to update the model parameters $(\theta_\beta,\theta_{\bf q})$ and $\theta_{\bm \xi}$ at the start of the $t$th period.
Since the training samples can be obtained by sliding the observation window in each time period, the model parameters can be updated sequentially. As a result, the optimized proactive policy can adapt to the dynamic popularity.

With the trained DNNs, the output of $\beta^t\left({\bf h}\right)$ and ${\bf q}^t\left({\bf h}\right)$ can be obtained in real-time via forward propagation at the start of the $t$th time period.

\section{Simulation Results}
In this section, we evaluate the performance of the proposed framework in the example system. To show the impact of dynamic popularity on the optimized policy with implicit popularity prediction, we use a real dataset for simulation.


The densities of the BSs and users are set as $\lambda_b=\lambda_u=5/(250^2\pi)$~m$^{-2}$. The transmit power is $P=30$ dBm. The total bandwidth is $W=20$ MHz. The data rate threshold is $R_0=2$ Mbps. The path loss exponent is $\alpha=3.7$ \cite{Simupara}.
The caching and bandwidth allocation policy is updated each day, and the observation window is set with length of $\tau=5$ days.

\subsection{Real Dataset and Sample Generation}
We consider a YouKu (a famous video on demand website in China) dataset, which consists of $239927$ requests for $110600$ files from $11205$ anonymized users  in a $2$~km$^2$ region during $86$ consecutive days from Aug. $28$th to Nov. $21$st, $2016$.
In the dataset, approximately $98 \%$  of the files are with less than $10$ requests in total, which are unnecessary to cache. Hence, the training samples are only generated from the requests for $2\%$ files in the dataset.
To generate sufficient samples for training the DNNs, we use the following repeat-sampling method.\footnote{We have also simulated using the popularity synthesized with the shot noise model widely-applied for caching \cite{SNMmode}, and found that the results are similar. Such a repeat-sampling is unnecessary for the synthesized dataset.}

Denote a record of the $f$th file as $[\hat p_f^{t};\hat p_f^{t-1},...,\hat p_f^{t-\tau}]$, where $\hat p_f^j$ $(j=t,...,t-\tau)$ is the popularity estimated\footnote{Recall that the popularity of a file is defined as the probability that the file is requested. We can only obtain an estimated popularity since only the number of requests for a file  can be recorded in a real dataset.} from dividing the number of requests for the $f$th file by the total number of requests for these $2\%$ files in the $j$th time period. We generate multiple records by sliding the observation window from the 2nd day to the $86$-th day, and then we can generate a set with $19558$ records for these files, denoted as ${\mathcal N}_\mathrm{r}$. When generating a sample for unsupervised learning (i.e., $(\hat {\bf p}^t,\hat {\bf h})$), we randomly select $F$ records from the set ${\mathcal N}_\mathrm{r}$.

To generate training and test samples, we first divide ${\mathcal N}_\mathrm{r}$ into two sets ${\mathcal N}_\mathrm{tr}$ and ${\mathcal N}_\mathrm{te}$, which respectively occupy $80\%$ and $20\%$ of all records. Then, $10000$ training samples and $100$ test samples are generated from ${\mathcal N}_\mathrm{tr}$ and ${\mathcal N}_\mathrm{te}$ by using the repeat-sampling method, respectively. We further divide the training samples into two sets respectively to train the DNNs and to tune the hyper-parameters, which respectively occupy $75\%$ and $25\%$ of all the training samples.

\subsection{Caching Performance}\label{Num_1}

We evaluate the SOP achieved by the proposed framework (with  legend ``Unsup") by comparing with another end-to-end strategy of learning to optimize using supervised learning (with  legend ``Sup") and the divide-and-conquer strategy of first-predict-then-optimize (with legend ``Preopt").

For the ``Sup" strategy, we use DNNs to approximate the functions $\beta^t({\bf h})$ and ${\bf q}^t({\bf h})$, which also obtains the optimized policy from the historical popularity in a single step. In order to train the DNNs with supervision, we first generate labels, each is obtained by solving ${\bf P0}$ with interior point method for a given estimate of ${\bf p}^t$, denoted as $(\beta^*(\hat {\bf p}^t),{\bf q}^*(\hat {\bf p}^t))$. Then, $(\hat {\bf h};\beta^*(\hat {\bf p}^t),{\bf q}^*(\hat {\bf p}^t))$ is the training sample for supervised learning. By generating labels in this way, the DNNs can also predict the popularity implicitly, since the policies for the $t$th time period are optimized at the start of the $t$th period using the estimated popularity at the period (i.e., $\hat {\bf p}^t$). After generating multiple training samples, we can use the empirical mean square error (MSE) between the output of the DNN with input $\hat {\bf h}$ and the expected output $(\beta^*(\hat {\bf p}^t),{\bf q}^*(\hat {\bf p}^t))$ (i.e., the labels) as the loss function to train the DNNs.

The ``Preopt" strategy first predicts the popularity for each file using a linear model \cite{WJJ2019APCC}, then  optimizes the caching and bandwidth allocation policy from ${\bf P0}$ with interior-point method by treating the predicted popularity as the true value.

Since the DNNs for ``Sup" and ``Preopt" can only be trained off-line, all the following results are obtained by the models that are well-trained in an off-line manner for a fair comparison.

To reflect the impact of implicit prediction embedded in proactive optimization,  we also show the performance achieved by the strategies using unsupervised and supervised learning with the future popularity of a file estimated from the future numbers of requests for the file known \emph{a priori}. Specifically, the training sample for ``Sup" is $(\hat {\bf p}^t;\beta^*(\hat {\bf p}^t),{\bf q}^*(\hat {\bf p}^t))$, i.e., the input of the DNNs is the estimate of future popularity rather than the historically estimated popularity. The optimal policy of  ``Unsup" is learned with the training samples of $\hat {\bf p}^t$ by using the Lagrangian function of problem  ${\bf P0}$ as the loss function.

Considering that the concerned policy only depends on the ranking of the file popularity, the file index is not a useful feature to train the DNNs for learning the proactive policy. To help the DNNs not to learn such a useless feature, we rank the samples for training the DNNs for ``Unsup" and ``Sup". For the methods with the estimated future popularity, the elements in each training sample are arranged in a descending order according to the estimated popularity of files. For the methods with the implicitly prediction, the elements in each training sample are arranged in a descending order according to the popularity in the ($t-1$)th period since the future popularity in the $t$th period in unknown \emph{a priori}.

We set cache size as $C=\frac{1}{10}F$. After fine tuning, the hyper-parameters for unsupervised learning are shown as follows. The DNNs $\tilde{\bf q}( {\bf h};\theta_{\bf q} )$, $ \tilde{\beta} ({\bf h};\theta_\beta)$, $\tilde{\xi}_c({\bf h};\theta_{ \xi_c} )$ and $\tilde{\bm \xi}_{f}({\bf h};\theta_{{\bm \xi}_{f}} )$ are with $3$, $1$, $1$ and $2$ hidden layers, where the number of nodes in each hidden layer is $[300,200,100]$, $[200]$, $[200]$ and $[200,100]$, respectively.
The hyper-parameters of the DNNs for the supervised learning are the same as  those in $\tilde{\bf q}( {\bf h};\theta_{\bf q} )$ and $ \tilde{\beta} ({\bf h};\theta_\beta)$ for unsupervised learning.
For all these DNNs, the learning rate is $0.1/(1+0.001i)$ in the $i$th iteration, the batch size is $32$, and the activation function for hidden layers is \texttt{ReLU} function. The number of epochs is $200$.

\begin{figure}[htbp]
    \centering
    \includegraphics[width=0.35\textwidth]{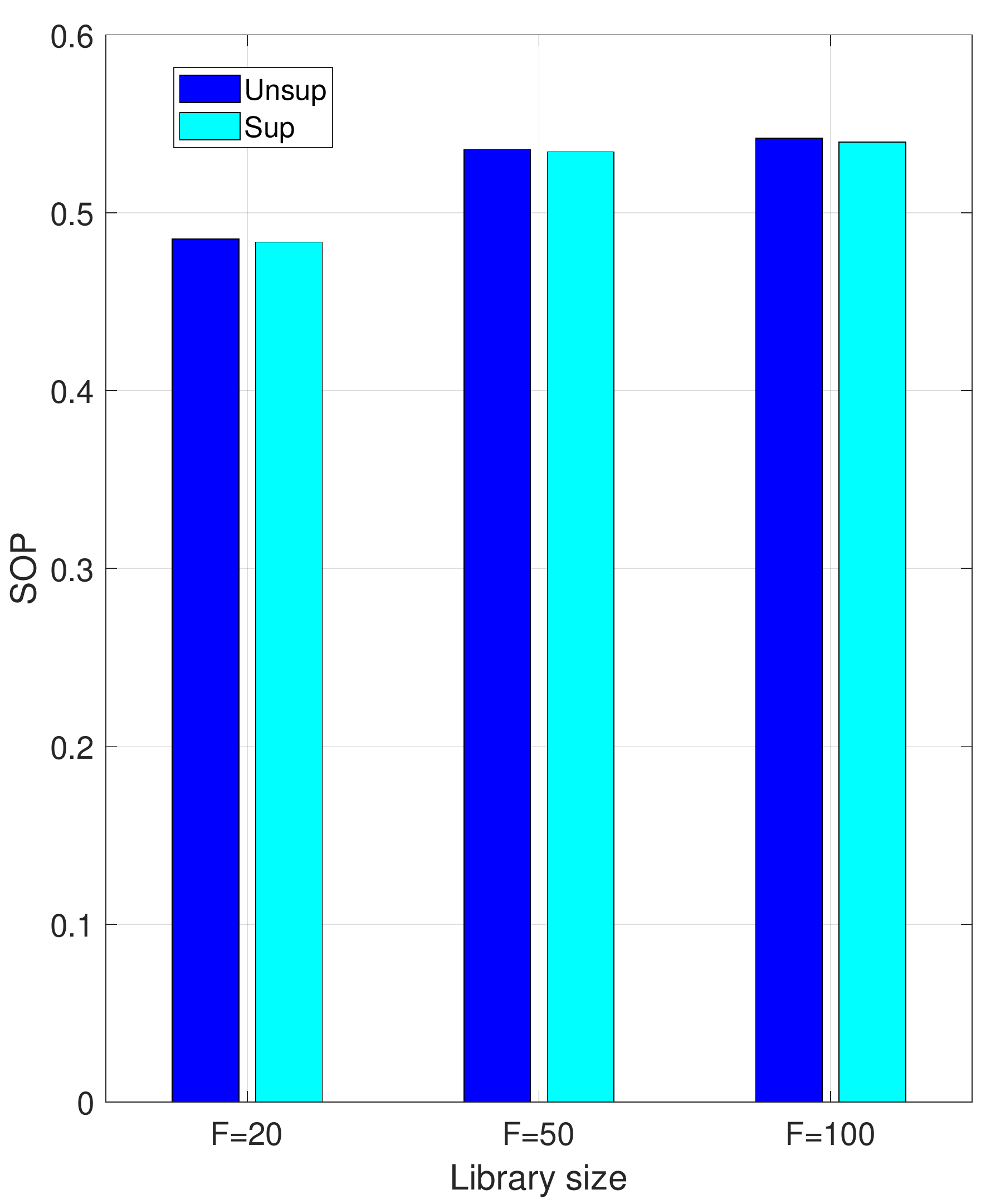}\vspace{-2mm}
    \caption{Caching performance comparison with estimated future popularity.}
    \label{SOP_Comp_opt}
\end{figure}
\begin{figure}[htbp]
    \centering
    \includegraphics[width=0.35\textwidth]{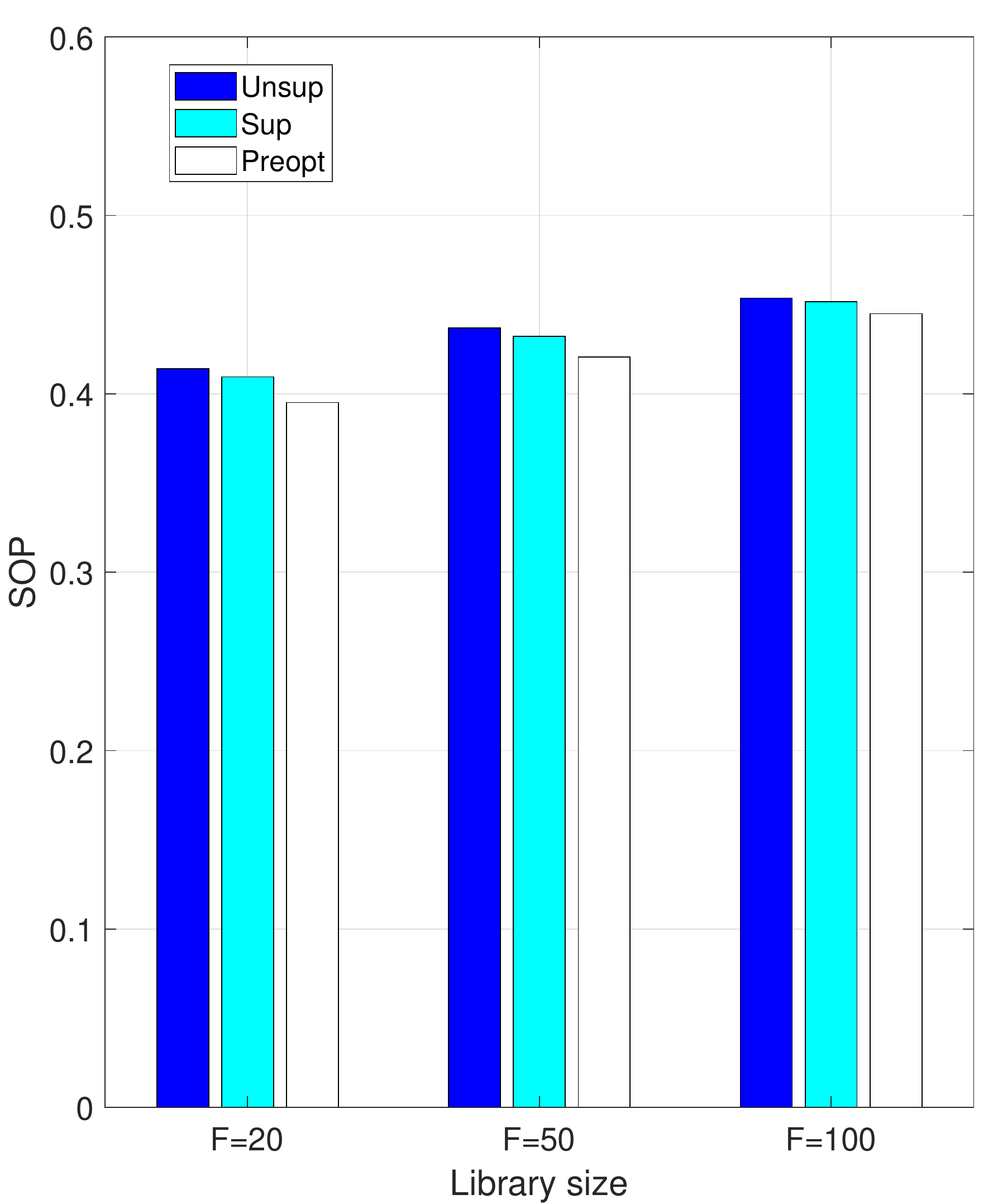}\vspace{-2mm}
    \caption{Caching performance comparison with predicted future popularity.}
    \label{SOP_Comp_pre}
\end{figure}

In Figs. \ref{SOP_Comp_opt} and \ref{SOP_Comp_pre}, we compare the SOP achieved by different strategies with the estimated future popularity and the predicted popularity on the test set, respectively. By comparing the results in the two figures, we can observe the performance loss caused by the prediction. From Fig. \ref{SOP_Comp_pre}, we can see that the proposed proactive framework with unsupervised learning is superior to the heuristic end-to-end strategy with supervised learning, and both strategies with implicit prediction perform better than the ``Preopt" strategy. Moreover, with the proposed framework with unsupervised learning, the complexity for generating the labels for supervised learning can be avoided, and only a single step is required for prediction and optimization.

\vspace{-2mm}

\section{Conclusions and Discussions}
In this paper, we introduced a framework to optimize proactive resource allocation policies, which can harness the unknown future information by leveraging the historical observations. We illustrated how to formulate and solve a proactive optimization problem by taking the proactive caching and bandwidth allocation problem as an example, where the content popularity is unknown \emph{a priori}.
Simulation results for the example problem with a real dataset validated that such an ``end-to-end'' strategy for prediction and optimization outperforms the ``first-predict-then-optimize" strategy and another ``end-to-end'' strategy using supervised learning. Different from the ``first-predict-then-optimize'' strategy, the proactive policy can be obtained in a single step with our framework. Since unsupervised learning was employed to find the solution, the proposed framework can optimize proactive policy without the high-complexity off-line phase of generating labels, and can adapt to the dynamic environment by updating the model parameters of the DNNs in an on-line manner. How to apply the proactive optimization framework to other wireless tasks deserves further investigation.


\vspace{-1.5mm}
\begin{appendices}
\numberwithin{equation}{section}
\section{Iteration Formulas}\label{App:Iter}
For notational simplicity, we let $\tilde{\bf q}$, $\tilde{\beta}$, $\tilde{\xi}_c$ and $\tilde{\bm \xi}_{f}$ denote $\tilde{\bf q}({\bf h};\theta_{\bf q}) $, ${\tilde \beta}({\bf h};\theta_\beta )$, $\tilde{ \xi}_c({\bf h};\theta_{\xi_c})$ and $\tilde{\bm \xi}_{f}( {\bf h};\theta_{{\bm \xi}_{f}} )$ in appendix, respectively.

Let $\mathcal B$ denote a batch of realizations of $\left({\bf p}^t,{\bf h}\right)$ and denote $|\mathcal B|$ as the number of the realizations. Then, the primal and dual variables are updated by
\begin{subequations}
\begin{equation}
\label{ite_q}
\begin{split}
\theta_{\bf q}^{i+1} \leftarrow  \theta_{\bf q}^{i} & + \frac{\delta}{|\mathcal B|} \sum_{({\bf h},{\bf p}^t) \in \mathcal B}  \nabla_{\theta_{\bf q}} \tilde{\bf q} \\ & \Big[\nabla_{ {\bf q}^t }  p_s\big({\bf p}^t, {\beta}^t,{\bf q}^t \big)\big|_{{ \beta}^t=\tilde{\beta},{\bf q}^t=\tilde{\bf q}} \!-\!\tilde{\bm \xi}_c\!-\!\tilde{\bm \xi}_{f}\Big],\nonumber
\end{split}
\end{equation}
\begin{equation}
\begin{split}
\label{ite_b}
\theta_{\beta}^{i+1} \leftarrow \theta_{\beta}^{i}&+\frac{\delta}{|\mathcal B|} \sum_{({\bf h},{\bf p}^t) \in \mathcal B} \nabla_{\theta_{\beta}} \tilde{\beta}  \nabla_{{ \beta}^t}p_s\big({\bf p}^t,{\beta}^t, {\bf q}^t \big) \big|_{{ \beta}^t=\tilde{\beta},{\bf q}^t=\tilde{\bf q}},\nonumber
\end{split}
\end{equation}
\begin{equation}
\begin{split}
\label{ite_xic}
&\theta_{\xi_c}^{i+1} \leftarrow \theta_{\xi_c}^{i}+\frac{\delta}{|\mathcal B|} \sum_{({\bf h},{\bf p}^t) \in \mathcal B} \nabla_{\theta_{\xi_c}} \tilde{\xi}_c  (\sum_{f=1}^{F}\tilde{q}_f-C ),\nonumber
\end{split}
\end{equation}
\begin{equation}
\begin{split}
\label{ite_xif}
&\theta_{{\bm \xi}_{f}}^{i+1} \leftarrow \theta_{{\bm \xi}_{f}}^{i}+\frac{\delta}{|\mathcal B|} \sum_{({\bf h},{\bf p}^t) \in \mathcal B}  \nabla_{\theta_{{\bm \xi}_{f}}} \tilde{\bm \xi}_{f}(\tilde{\bf q}-1),\nonumber
\end{split}
\end{equation}
\end{subequations}
where $\nabla_{\bf x}{\bf y}=[(\nabla_{\bf x}y_1),...,(\nabla_{\bf x}y_m)]$ denotes the transpose of Jacobian matrix, $\nabla_{\bf x}y=[\frac{\partial{y}}{\partial x_1},...,\frac{\partial{y}}{\partial x_n}]^T$ denotes the gradient, $\tilde{\bm \xi}_c=[\tilde{\xi}_c,...,\tilde{\xi}_c]$ is a $F $-dimension vector,
$\delta$ is learning rate, and $(\cdot)^T$ denotes the transpose of a vector.

According to the expression of SOP in \eqref{ps_ini}, the elements of the  gradients  $\nabla_{{\bf q}^t} p_s({\bf p}^t,{\beta}^t,$$ {\bf q}^t) $ and   $\nabla_{{\beta}^t}p_s({\bf p}^t,{\beta}^t,{\bf q}^t) $ can be computed as,
\begin{subequations}
\begin{equation}
\label{grad_q}
\begin{split}
\frac {\partial p_s  ({\bf p}^t,{\beta}^t, \tilde{\bf q}^t)}{ \partial {q_f^t}}= \frac{ K p_{\rm a}p_f^t{{\beta}^t} (\gamma_{0,{\beta}}^t)^{\frac{2}{\alpha}}} {\kappa_f^2 },\nonumber
\end{split}
\end{equation}
\begin{equation}
\begin{split}
\label{grad_b}
&\nabla_{{\beta}^t } p_s({\bf p}^t,{\beta}^t, {\bf q}^t)  =
  -\sum_{f=1}^{F} \Big\{\frac{p_{\rm a} p_f^tq_f^t}{\kappa_f^2 } \Big[q_f^t \big(Z_{1,\gamma_{0,{\beta}}^t}+
{\beta^t} \frac{ d Z_{1,\gamma_{0,{\beta}}^t}}{d {\beta}^t}  \big)
\\ &+K(1-q_f^t)(\gamma_{0,{\beta}}^t)^{\frac{2}{\alpha}}\big(1 +\frac{2}{\alpha}{\beta^t}(\gamma_{0,{\beta}}^t)^{-1}\frac{d \gamma_{0,{\beta}}^t}{d {\beta}^t } \big)\Big]\Big\},\nonumber
\end{split}
\end{equation}
\end{subequations}
where  $\kappa_f= q_f^t + p_{\rm a} {\beta^t}\big(q_f^t Z_{1,\gamma_{0,\beta}^t}+K(1-q_f^t)(\gamma_{0,\beta}^t)^{\frac{2}{\alpha}} \big) $, $\frac{d Z_{1,\gamma_{0,\beta}^t } } {d \beta^t} $\\ $= \frac{2}{\alpha} \frac{d \gamma_{0,\beta}^t} { d \beta^t}$ $ (( \gamma_{0,\beta}^t)^{\frac{2}{\alpha}-1} \int_{(\gamma_{0,\beta}^t)^{-\frac{2}{\alpha}}}^\infty \frac{1}{1+x^{\frac{\alpha} {2}}} \,dx+\frac{(\gamma_{0,\beta}^t)^{-1} } {1+(\gamma_{0,\beta}^t)^{-1}} ) $, \\  and $\frac{d \gamma_{0,\beta}^t}{ d \beta^t} \approx -\frac{R_0 }{ ({\beta^t})^2 W}(1+1.28\frac{\lambda_u}{\lambda_b})  2^{ \frac {R_0 } { { \beta^t} W}(1+1.28\frac{\lambda_u}{\lambda_b})  } \ln2 $.

\end{appendices}

\bibliographystyle{IEEEtran}
\bibliography{WJJ}
\end{document}